# Correlations between X-rays, Visible Light and Drive-Beam Energy Loss Observed in Plasma Wakefield Acceleration Experiments at FACET-II

Chaojie Zhang[1], Doug Storey[2], Pablo San Miguel Claveria[3,4], Zan Nie[1], Ken A. Marsh[1], Warren B. Mori[1,5], Erik Adli[6], Weiming An[7,8], Robert Ariniello[2], Gevy J. Cao[6], Christine Clark[2], Sebastien Corde[3], Thamine Dalichaouch[5], Christopher E. Doss[9], Claudio Emma[2], Henrik Ekerfelt[2], Elias Gerstmayr[2,10], Spencer Gessner[2], Claire Hansel[9], Alexander Knetsch[2,3], Valentina Lee[9], Fei Li[5,11], Mike Litos[9], Brendan O'Shea[2], Glen White[2], Gerry Yocky[2], Viktoriia Zakharova[3], Mark Hogan[2] and Chan Joshi[1]

[1]Department of Electrical and Computer Engineering, University of California Los Angeles, Los Angeles, CA 90095, United States of America
[2]SLAC National Accelerator Laboratory, Menlo Park, CA 94025, United States of America
[3]LOA, ENSTA Paris, CNRS, Ecole Polytechnique, Institut Polytechnique de Paris, 91762 Palaiseau, France
[4]GoLP/Instituto de Plasmas e Fusao Nuclear, Instituto Superior Tecnico, Universidade de Lisboa, 1049-001 Lisboa, Portugal
[5]Department of Physics and Astronomy, University of California Los Angeles, Los Angeles, CA 90095, United States of America
[6]Department of Physics, University of Oslo, Oslo, 0316, Norway
[7]Department of Astronomy, Beijing Normal University, Beijing 100875, China
[8]Institute for Frontiers in Astronomy and Astrophysics, Beijing Normal University, Beijing 102206, China
[9]Department of Physics, Center for Integrated Plasma Studies, University of Colorado Boulder, Boulder, CO 80309, United States of America
[10]Stanford Pulse Institute, Stanford University, Menlo Park, CA 94305, United States of America
[11]Department of Engineering Physics, Tsinghua University, Beijing 100084, China

**Abstract**

This study documents several correlations observed during the first run of the plasma wakefield acceleration experiment E300 conducted at FACET-II, using a single drive electron bunch. The established correlations include those between the measured maximum energy loss of the drive electron beam and the integrated betatron x-ray signal, the calculated total beam energy deposited in the plasma and the integrated x-ray signal, among three visible light emission measuring cameras, and between the visible plasma light and x-ray signal. The integrated x-ray signal correlates almost linearly with both the maximum energy loss of the drive beam and the energy deposited into the plasma, demonstrating its usability as a measure of energy transfer from the drive beam to the plasma. Visible plasma light is found to be a useful indicator of the presence of wake at three locations that overall are two meters apart. Despite the complex dynamics and vastly different timescales, the x-ray radiation from the drive bunch and visible light emission from the plasma may prove to be effective non-invasive diagnostics for monitoring the energy transfer from the beam to the plasma in future high-repetition-rate experiments.

**1.1 Introduction**

Plasma wakefield acceleration (PWFA) affords a significant advance in particle acceleration technology, by offering orders of magnitude larger acceleration rates [1] and high energy transfer efficiency from the drive bunch (beam) to the accelerating trailing bunch (beam) thereby holding the promise of more compact and cost-efficient next-generation high-energy lepton accelerators [2] and beam-driven light sources (e.g., free electron lasers) [3–5]. Key aspects of plasma wakefield acceleration are depicted in Fig. 1. Within a PWFA, an intense highly relativistic charged particle bunch, known as the 'drive bunch' or 'driver,' expels plasma electrons (in a pre-ionized plasma or plasma produced by transverse electric field of the driver itself) from their initial positions, creating oscillations (wake). If the driver bunch is intense enough the wake formed is in the nonlinear-blowout regime where the wake structure comprises an ion column surrounded by a plasma electron sheath and moves at the speed of the driver. As shown by the red lineout in Fig. 1, a strong longitudinal electric field exists inside the wake due to the charge separation of the ions and the electrons that can accelerate an appropriately positioned 'trailing bunch' of particles to high energies over remarkably short distances [6].

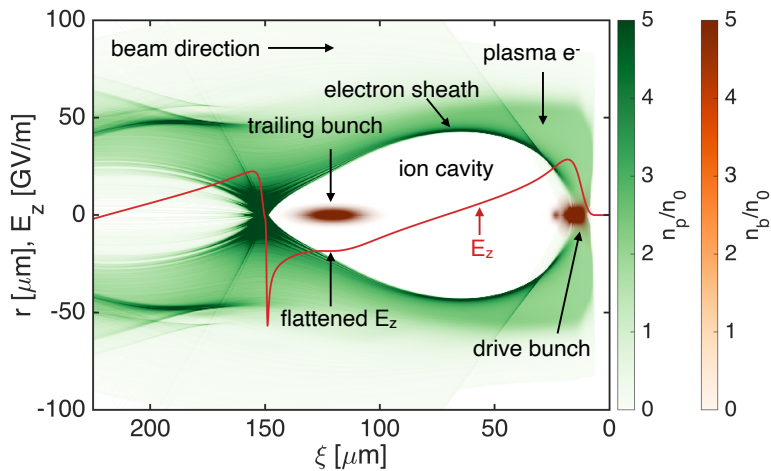

*Fig. 1. Schematic of two-bunch plasma wakefield acceleration. An intense beam (drive bunch) ionizes lithium vapor to create plasma (illustrated by the green colormap) and excites a wake, characterized by an ion cavity enclosed by a plasma electron sheath. The red line indicates the*

*longitudinal field within the wake (along the axis). A trailing bunch, placed in the accelerating phase (negative $E_z$ region) with an appropriate amount of charge can flatten the accelerating field, enabling rapid energy gain while preserving its absolute energy spread. Notice that in the configuration illustrated here, the magnitude of the flattened (uniform) electric field is 18 GeV/m.*

The drive bunch loses energy to the plasma as it excites the wake. The plasma response depends primarily on the current and transverse size of the driver, leading to the formation of a mostly non-evolving wake, whose topology and therefore the fields do not change with propagation distance in the plasma. This remains true until the driver begins to be significantly depleted of its energy. In such a non-evolving wake the energy extraction gradient from the wake to the trailing bunch remains more or less the same over a unit length of plasma. On the other hand, the energy lost to the plasma wake by the driver increases by either increasing the wake amplitude or the distance over which the wake is produced. In either case, the net drive-to-trailing bunch energy transfer efficiency is maximized when the driver is nearly fully depleted of its energy and thus no longer able to sustain the wake.

One of the objectives of the E300 experiment campaign is to optimize the energy extraction efficiency from a 10 GeV electron drive bunch to the plasma wake at the FACET II facility. FACET-II, situated at the SLAC National Accelerator Laboratory, is designed to provide ultra-relativistic $\gamma > 10^3$ electron bunches with ultra-high peak currents (30-100 kA) for advanced accelerator research [7]. As the successor to the original FACET, this newly operational facility is designed to provide versatile beam configurations, including single- and double-bunch modes, with a nominal energy of 10 GeV and an emittance of better than 20 microns. These features are instrumental for validating key PWFA concepts, such as driver energy depletion, efficient energy transfer from driver to trailing bunch, and maintaining both the trailing bunch's minimal energy spread and its emittance through optimal wake loading and beam matching [6].

The data presented here was obtained during the commissioning phase of the FACET II facility in late 2022. During the ramp-up stage, the FACET-II facility was configured to generate a single-bunch capable of delivering high charge (>1 nC), highly compressed (peak current >50 kA) bunches. Using this single-bunch configuration, we have conducted PWFA experiments with a plasma produced by high-field ionization by the drive beam itself of a 4-meter-long column of hydrogen (or helium) gas. Noteworthy findings from these experiments include direct observation of the onset of near-total energy depletion of some of the electrons in the initial 10 GeV drive bunch, a clear dependence of energy loss on plasma density, an impressive energy transfer efficiency of up to 60% from (unoptimized) driver to wake, and the acceleration of some of the electrons contained in the tail of the driver to multi-GeV energies. These results were reported in our recent publication [8].

**1.2 Correlations discovered In Past PWFA Experiments and New Correlations Observed**

In addition to deterministic diagnostics, correlations among other readily quantifiable diagnostics are very important because FACET II- like many other accelerator facilities, is able to operate at up to 30 Hz. At such high repetition rates, it is practically impossible to evaluate the significance of the data in real time. The data must be saved so that it can be retrieved at a later time/date. The issue, that will become ever more important as the FACET-II experiments transition from the current 1 Hz to higher repetition rates, is which sets of data should be saved (revisited) for further deeper analysis. This is where correlation plots can come in useful.

In the past we have used three readily quantifiable diagnostics on FACET and an earlier facility FFTB that have proved to be extremely useful and we will continue to use them on FACET II. The first is the magnitude of the signal on a diode that measured the coherent component of the transition radiation (CTR) produced by the fully compressed bunch as it travelled through optical transition radiation (OTR) screens- thin metal foils- placed upstream and downstream of the plasma as a measure of the peak current of the single drive bunch [9]. CTR signal is produced when frequency spectrum of the OTR radiation contains bright peaks in the visible part of the spectrum which is indicative of one or more spikes of high peak current whose duration is on the order the wavelength of visible light. We observed that larger the CTR signal the larger was the peak energy lost by the beam.

A second diagnostic was the measurement of excess charge over and above the charge contained in the drive and trailing bunches. This excess charge was measured by comparing the charge entering the plasma with that exiting the plasma using toroidal transformers placed on either side of the plasma. Excess charge is indicative of injection and acceleration of some of the charge produced during the beam ionization into the wake and therefore exiting with the beam electrons, albeit with a different energy. During the first round of PWFA experiments at the FFTB facility direct evidence of these injected electrons was seen on a secondary threshold Cherenkov spectrometer as well as excess charge measured by the downstream toroid [10]. This correlation was observed again on FACET where the distributed injection of the excess charge into the wake was acted as dark current that beam loaded the Rubidium wake [11]. This so-called ionization injection could be used to generate a low emittance beam to give a higher brightness secondary beam for plasma-FEL applications. In laser wakefield acceleration experiments ionization injection both reduces the threshold for plasma electron self-injection [12] as well as makes the experimental outcome more reproducible.

The third was an imaging diagnostic that gave information about the energy dissipation time of the wake. This information was obtained by measuring transverse density fluctuations left in the expanding plasma long after the passage of the drive beam by a shadowgraphy technique. This showed that at 1 Hz repetition rate, the energy left behind in the wake by either a single drive beam or by a drive-trailing bunch pair is not fully dissipated for up to 10 microseconds [13] in a transversely confined gas, strongly suggesting that for high rep-rate operation the wake may need to be generated in a transversely flowing gas. Correlation plots of dissipation times for different gas flow rates and for different gases are needed to determine a reasonable repetition rate for a given gas.

In this paper, we document several correlations between energy gain/loss diagnostics and other readily quantifiable diagnostics. Specifically, the experimental data from the first FACET-II run revealed a linear correlation between both the maximum energy loss and total energy deposited by the drive bunch and the integrated betatron x-ray radiation. This suggests that, once properly calibrated, the betatron x-ray signal may serve as- at least a qualitative- indicator of beam energy transfer to the wake at least until the wake is still in the non-evolving stage. Remarkably, a recent simulation study performed in the context of FACET-II has also found clear correlations between the integrated betatron yield and the emittance preservation of the accelerated beam [14]. We also observed correlations among multiple cameras that monitor the visible plasma light at different locations. These correlations give an estimate of the length of the wake and the differences between them gives additional indication of the onset and termination of the wake by pump depletion. Another surprising correlation exists between the visible plasma light and the betatron x-ray signal, because the two processes are expected to occur on very different timescales.

**2 Experimental Setup, Relevant Diagnostics and Examples of Data**

In a previous publication [15], we detailed the beamline configuration of FACET-II, the experimental setup around the interaction point (IP) area, and the diagnostics. Here, we discuss the setup and relevant diagnostics pertinent to the results reported in this paper (Fig. 2) for the reader's convenience.

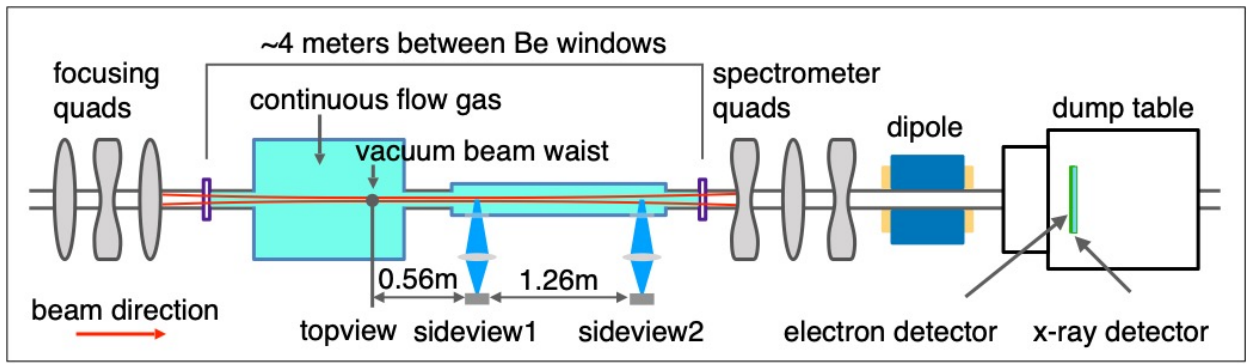

*Fig. 2. Sketch of the experimental setup. The 10 GeV single-bunch driver propagates from left to right, being focused by a set of quadrupoles into a gas filled region, which is bounded by two beryllium windows (with beam-drilled holes). The gas pressure within this region is adjustable, maintained at less than 5 Torr, constrained by the capabilities of the differential pumping system (not shown in the sketch). The post-interaction electron bunch is analyzed by the downstream diagnostics, which include an imaging spectrometer (composed of tunable quadrupoles, a vertically dispersing dipole magnet, and a large field of view (LFOV) Gadolinium Orthosilicate based scintillator screen coupled with a camera) and x-ray detectors (GAMMA1, detailed in ref. [16]). Three cameras (topview, sideview1, and sideview2) capture the time- and spectral-integrated plasma light at various locations. The distance between these cameras is indicated on the figure.*

As previously mentioned, FACET-II currently operates in a single-bunch configuration, delivering 10 GeV high-brightness electron bunches with peak currents exceeding 50 kA. These peak currents are a factor of 2-3 higher than in previous experiments on FACET and at FFTB and thus have allowed ionization of higher ionization potential (IP) gases (such as hydrogen molecules $H_2$ with IP 15.4 eV and He atoms with IP 24.6 eV) compared to previously used Lithium (IP 5.4 eV). In the present experimental results reported here, the compressed bunch is focused by a set of quadrupoles into an approximately 4-meter-long continuous flow gas region, achieving a root mean square (rms) spot size of ≳20 µm. This region is confined by two 50 µm thick beryllium windows with approximately 200 µm diameter holes drilled *in situ* by the electron bunch itself. The pressure of the hydrogen gas flow is adjustable from a few mTorr up to 5 Torr and can be maintained within 0.5% precision on the time scale of a day. A four-stage differential pumping system on the upstream side has been implemented to isolate the high-pressure gas region from the high vacuum conditions upstream (e.g., better than $10^{-9}$ Torr at the x-band deflecting cavity). Similarly, a two-stage differential pumping system on the downstream side has been implemented to get better than $10^{-6}$ Torr pressure to reduce beam scattering as the beam propagates to the end of the beamline.

In the experiment, the vacuum focus of the beam was placed at position of the topview camera and the vacuum beta of the beam is about 50 cm. The spot size of the beam is related to the beta function $\beta$ through $\sigma = \sqrt{\epsilon\beta}$, where $\epsilon$ is the geometric emittance of the bunch. As the electron bunch approaches the vacuum focus region, it initiates ionization induced by the transverse electric field of the beam within the hydrogen gas (field-induced ionization) even before the beam has reached the vacuum focus position which is close to the center of the field of view of the top view camera, one of three cameras that look at the plasma light emission. For bunches with peak current above ~30 kA (exceeding the ionization threshold of hydrogen molecules for a $\sigma_r$ =30 µm Gaussian bunch) the hydrogen gas is ionized earlier in the risetime of the beam current over a length of several meters far greater than the vacuum beta of the beam suggesting that the beam is both focused and guided by the plasma that it has created. Plasma formation is immediately followed by wake formation. In transversely displacing the plasma electrons the beam electrons lose kinetic energy which mostly appears as the wake potential. The majority of the bunch loses energy to the wake; however, the very back of the bunch can gain energy if the plasma density is high enough.

After interaction with the plasma, the electron bunch is analyzed by downstream diagnostics that measure its charge, energy spectrum, and emittance, as well as the total x-ray signal. Surrounding the gas region, there are two additional cameras to the already mentioned topview camera- sideview 1 and sideview 2 in Fig.2- that also capture the time- and spectrally-integrated plasma light emission (within the visible range, i.e., 400-700 nm). All these diagnostics are capable of operating on a single-shot basis [17] and can in principle be operated at up to 30 Hz.

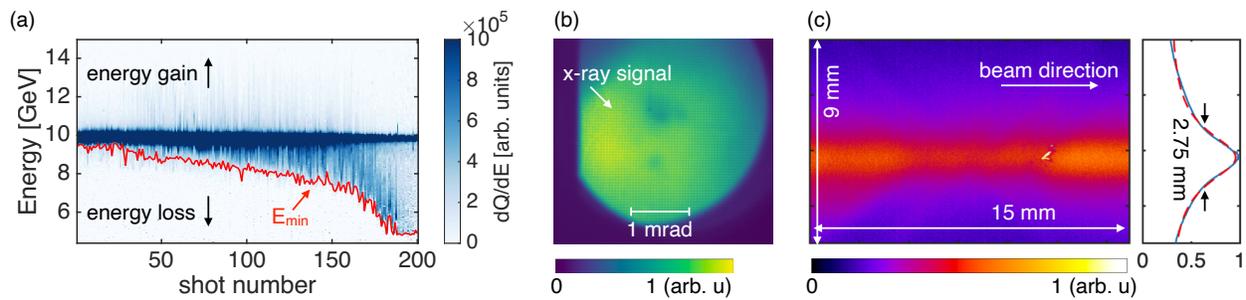

*Fig. 3. Representative data samples. (a) Linearized electron energy spectra of a set of 200 shots, sorted by increasing integrated x-ray signal on the screen shown in (b). The red line marks the minimum energy ($E_{min}$) detected by the spectrometer (corresponding to the maximum energy loss $E_{max\ loss}$). (b) A typical x-ray signal as recorded by the GAMMA1 detector. Note that the hole in the middle and the other smaller hole underneath are not features of the betatron radiation, but are due to regions of decreased light output of the detector due to radiation related damage, and persist unchanged shot-to-shot. The defects cause the local apparent yield to be ~15% smaller. (c) An illustrative image of the plasma light emission captured by the sideview1 camera with an exposure time of 50 µs. The adjacent plot displays the integrated signal along the beam direction (solid blue line), along with a Lorentzian fit (dashed red line). The Lorentzian fit has a full width at half maxima of 2.75 mm and 80% of it is within the transverse field of view of the camera.*

In Fig. 3, we present examples of data acquired by the energy spectrometer, the integrated betatron emission screen and visible emission recorded by the sideview1 camera. Fig. 3(a) displays the linearized energy spectra of the drive bunch for a dataset of 200 consecutive shots, taken with hydrogen gas at 1.0 Torr. Since the data was taken during the commissioning phase of the facility, with not-yet optimized beam shot-to-shot stability, it resulted in a fluctuating beam-plasma interaction strengths within the same dataset. The LFOV recorded the dispersed energy spectrum of the electron bunch after it has both produced and interacted with the plasma for each shot. The raw spectrum was first integrated along the undispersed direction and subsequently linearized along the dispersed direction, using the pre-calibrated dispersion curve of the dipole magnet. The dataset is sorted in ascending order according to the x-ray signal. In all shots, the prominent peak at 10 GeV represents the charge that either did not interact with the plasma (e.g., electrons preceding the ionization front) or did not undergo significant energy change (e.g., electrons at the zero-field phase of the wakefield). A prior publication has established that this 'non-participating' charge comprises up to 30% of the total charge for all gas pressures used, indicating >30 kA peak current at the front of the drive bunch [8]. The red line in Fig. 3(a) marks the minimum energy ($E_{min}$), corresponding to the maximum energy loss ($E_{max\ loss} = 10 - E_{min}$) by the drive bunch, except for the last 25 shots where the energy loss exceeded the spectrometer's measurement limit (≳5 GeV) in this configuration.

Fig. 3(b) presents a representative image of the betatron x-ray signal captured by the "GAMMA1" detector, which records the spatial profile of the x-ray signal without resolving the spectrum. The setup and calibration details of this detector are provided in [16]. In the particular shot displayed, the drive bunch experienced a maximum energy loss of approximately 5 GeV. Although the left and top edges of this screen are partially shadowed and about 15% of the data is missing, the x-ray intensity profile appears round with a full width at half maximum (FWHM) of about 3 mrad. The betatron emission aperture angle is $\theta \approx (1 + K)/\gamma$ where $K = k_p r_\beta \sqrt{\gamma/2}$ is the equivalent of the undulator strength parameter in synchrotrons, $r_\beta$ is the betatron oscillation magnitude and $\gamma$ is the Lorentz factor of the electrons. The upper estimated limit of the aperture angle is estimated to be ~1 mrad by using an energy of 5 GeV (average energy loss of 5 GeV), an emittance of 30 μm, and $r_\beta$ approximated by the matched spot size of 3.5 μm (corresponding to $K \sim 8.3$). The observed x-ray emission angle, being three times larger, suggests that the bunch is not matched, which is further supported by the significant (more than an order of magnitude) increase in its emittance [15].This emittance growth would increase the average matched size of the bunch, increasing $r_\beta$ and thereby the angle.

In Fig. 3(c), a representative image of the plasma light captured by the sideview1 camera is shown. For this run, no color filter was used, allowing the camera to record plasma emissions across the complete wavelength range of approximately 400-700 nm. The camera's exposure time was set to 50 µs, far longer than the lifetime of the wakes [13,18,19]. The image shows a stripe a few millimeters wide that indicates the plasma width as the beam travels from left to right across the field of view. The line plot adjacent to the image shows the integrated signal over a 2.5 mm region (within the two dashed white lines) along the beam direction, which is well fitted by a Lorentzian function with a background. The Lorentzian part of the signal is likely due to the direct imaging of the plasma emission, while the background may be explained due to multiple reflections of the plasma emission light within the chamber. The plasma light emission has a FWHM core of about 2.7 mm. The transverse field of view of the camera collects about 80% of the Lorentzian part of the signal. That is to say the light comes from the expanding plasma core is likely to be the main factor responsible for correlations observed with x-ray emission that will be described later.

**3 Correlations identified in PWFA experiments**

In this section, we discuss the correlations identified between these observables, which have proven to be useful in enhancing our understanding of the nonlinear beam-plasma interactions. We note that for these correlations we do not model the details of the betatron x-ray emission nor of the visible light emission by the plasma. The reason is that both the betatron and the integrated visible light data is likely to be experiment specific (namely, they depend on beam and diagnostics configurations).

During the experiment for a given setting of the beam parameters one or more sets of 200 shots are recorded. The correlation plots such as those shown below can be produced within minutes of the end of the certain beam configuration setting and can be recorded as a measure of likely usefulness of the data that needs further analysis.

For example, one objective of the E300 experiment at FACET-II is to demonstrate high energy transfer efficiency from the drive to the trailing bunch. This efficiency is the product of the drive bunch to the wake energy transfer $\eta_{D-W}$ and the energy transfer from the wake to the trailing bunch $\eta_{W-T}$. In the case of a nonevolving wake the product of the two efficiencies can be estimated from a measurement of partial energy loss of the drive bunch and energy gain by the trailing bunch [20,21]. For an experimental demonstration of a single stage of a multi-stage PWFA-based linear collider, the energy of the drive bunch must be fully depleted in producing the wake. Ideally, energy depletion happens when all the particles in the beam are decelerated by the wake at the same rate and therefore the minimum energy recorded by the spectrometer approaches zero for all the particles in the beam, or alternatively, when the maximum energy loss ($E_{max\,loss}$) approximates the drive bunch's initial energy. However, this situation requires a specially shaped current profile of the electron beam [22–24]. Since in the present work the current profile of the bunch is not optimized, different longitudinal slices of the bunch loose energy at a different rate. This means while electrons in a portion of the beam may very well be pump depleted, those in other longitudinal slices will not be fully depleted. Indeed, if the beam has a long tail, then the electrons in the tail may well be in the accelerating phase of the wake and therefore gain energy. There is also a practical problem to making the ideal measurement. The post-interaction beamline (from the exit of the plasma to the energy spectrometer screen, may only be optimized to focus and transport electrons only within a specific energy range. This is less problematic for lower-energy electron drive beams; for example, recent studies have demonstrated near-total energy loss of a 500-MeV drive bunch to the wake in discharge plasmas inside a 15-cm-long capillary [25]. For higher-energy drive bunches, like the 10 GeV used at FACET-II, this method is not feasible because it would require transporting electrons that have an excessively broad energy range. For instance, a two-bunch experiment aimed at doubling the trailing bunch's energy while depleting the drive bunch would require a beamline capable of handling electrons with energies ranging from 0 to 20 GeV, which is not possible within a single optics configuration. In addition to maximum energy loss, the total energy deposited into the plasma by the drive bunch is also important, as it relates to the efficiency of energy transfer from the drive to the trailing bunch. Assuming that energy dissipation through other mechanisms such as betatron x-ray emission and ionization of the neutral gas by the drive bunch is negligible, we calculate the energy deposited into the plasma as the difference between the initial and post-interaction energies of the drive bunch. However, due to some low-energy electrons not being captured and transported to the spectrometer imaging screens as previously explained, this method can encounter significant uncertainties. Consequently, both scenarios necessitate the development of alternative methods that can be calibrated to rapidly give the maximum energy loss and the energy deposited into the plasma. This is particularly crucial when the imaging energy of the spectrometer optics is set to a few GeV to ascertain the emittance of decelerated electrons within a narrow energy range around the imaging energy, employing the butterfly technique where the width of beam in the undispersed direction reaches the minimum at the focusing energy and increases at non-focusing energies [15].

**3.1 Correlation between betatron x-ray signal and maximum energy loss and total energy lost by the beam**

We have discovered that the total incoherent but directional (betatron) x-ray emission is an effective indicator of the initially 10 GeV beam's maximum energy loss ($E_{max\,loss}$) within the spectrometer setting limit (up to 5 GeV in this case). The maximum energy loss is determined by the product of the maximum decelerating field experienced by a slice of the beam and the plasma length at a given density. Conversely, the betatron x-ray signal not only depends on these parameters (wakefield strength, and length over which the wake exists) but also on the beam's energy and the particles' displacement from the propagation axis within the beam. The longitudinal phase space of the beam is continuously evolving as the beam slices lose energy by different amount to the wake. In principle a 3D moving window PIC simulation allows one to model the electron trajectories and these can be coupled to a radiation code to determine the x-ray yield [26,27].

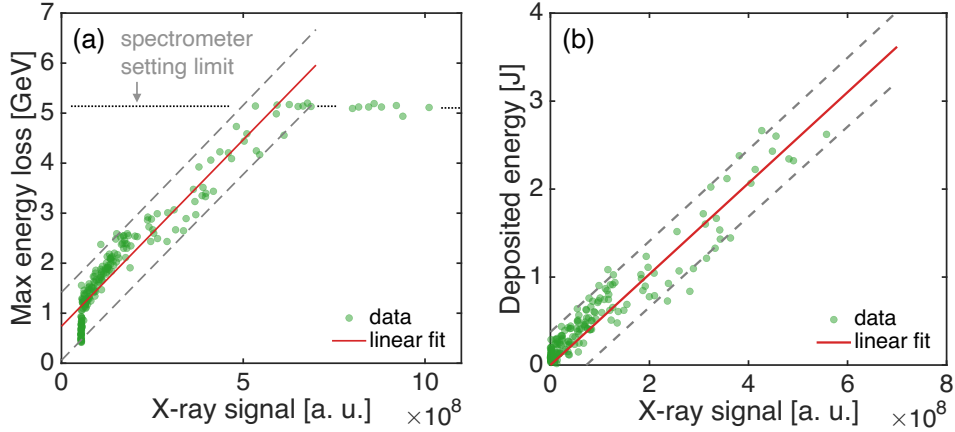

*Fig. 4. Correlations between the maximum energy loss of the drive bunch (a), the energy lost by the beam while traversing the plasma (b), and integrated betatron x-ray signal. In both (a) and (b), the data (green dots) exhibit a strong correlation, as demonstrated by the linear fit (red line), with the two dashed lines indicating the 95% prediction interval of the fit. Note that in (a), data points with a maximum energy loss remaining at ~5 GeV, as the x-ray signal increases beyond $8 \times 10^8$, are due to the spectrometer's setting limit for this dataset and have been excluded from the linear fit.*

In our experiment, despite the complex dependencies of the x-ray yield on beam, wake, and plasma parameters, we observed linear correlations between the maximum energy loss and the x-ray signal, as well as the total beam energy deposited into the plasma and the x-ray signal, as shown in Fig. 4(a) and (b) respectively for shots that the entire bunch charge remains visible on the profile monitor. Since the beam electrons are relativistic, beam-plasma electron collisions are a not significant mechanism of energy transfer. The energy is transferred in a collisionless process mediated by the transverse space charge field of the highly relativistic beam. In this process the plasma electrons are pushed outward by the repulsive force of the drive beam electrons. While most of them are eventually captured in a thin sheath by the attractive force of the plasma ions, some may escape. Here we assume that this is a small effect and all the energy lost by the beam electrons is transferred to the wake. The dataset analyzed here is the same as that shown in Fig. 3(a) and was acquired at a static hydrogen pressure of 1.0 Torr. The spectrally, and spatially integrated x-ray signal is calculated by summing the total counts for each shot.

In Fig. 4(a), there are data points for which the x-ray signal continues to increase (up to $10^9$ counts) even though, the maximum energy loss is capped at 5 GeV due to the spectrometer setting limit in this configuration. If projected onto the established linear correlation, these points would suggest a maximum energy loss of 8 GeV and a deposited energy of around 5 J. However, it should be noted that extrapolating the linear correlation must be verified with spectrometer settings that can measure the spectrum over an extended range to confirm its validity. We also note that the correlation almost certainly breaks down as a large fraction of the beam electrons are pump depleted since they can no longer emit betatron x-rays so eventually the x-ray emission must also begin to show saturation. What is clear that these correlation plots are experiment specific and over the range of measurement of both the variables, provide a playbook for rapid scanning of beam parameters during the experiment.

**3.2 Correlation among three different visible light cameras**

In addition to the x-ray diagnostic, we also used cameras to measure the visible light emitted by the plasma transversely at multiple locations around the interaction point (IP) area (see the topview and sideview cameras depicted in Fig. 2). Modeling light emission from a plasma with a set initial density and temperature is already a complex problem. In our experiments, this complexity is compounded as the plasma's peculiar initial state that must take into account plasma formation by beam field-ionization, energy deposition by the beam into the wake and dissipation processes including transverse shock wave generation [28] and is largely unknown (not diagnosed directly). Therefore, here we do not intend to establish a quantitative model to describe it but to show that despite the involved physics, the spectral- and time-integrated but spatially localized plasma light provides useful information about the beam-plasma interaction.

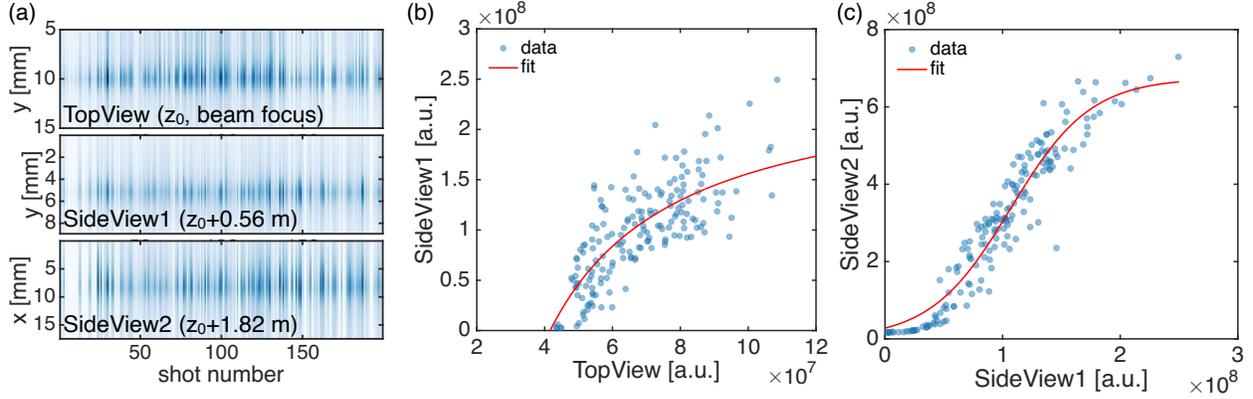

*Fig. 5. Correlations among different visible light diagnostics for 200 consecutive shots obtained at 10 Hz. (a) displays plasma light captured by the topview, sideview1, and sideview2 cameras, with each column corresponding to an individual shot and summed along the beam propagation direction. (b) illustrates the plasma light signal recorded by the sideview1 camera plotted against the signal from the topview camera. (c) shows the correlation between the signals from the two sideview cameras. The red lines in (b) and (c) are the best fit to the data using a power function and a logistic function, respectively, to show the nonlinear trend of the data.*

We first present correlations among the plasma light recorded by three cameras situated at different locations, which provides information of the plasma length over which the wake is excited. Then, we examine the correlation between the plasma light and the x-ray signal in the next section. Figure 5(a) displays plasma light collected by these cameras, with each column representing an individual shot summed over the beam propagation direction, and the dataset is presented in the order the data was acquired. The dataset was taken at 1.0 Torr. The top panel illustrates the light captured by the topview camera at the vacuum beam waist location, while the other two panels show the light collected by the sideview cameras, located 0.56 and 1.82 meters downstream, respectively. Correlations between these measurements are shown in Fig. 5(b) and (c), with the red lines showing the best fit using a power function and a logistic function, respectively, to indicate the nonlinear trend of the correlations.

In Fig. 5(b), the plasma light signal from the sideview1 camera is plotted as a function of the topview camera signal. Notably, the topview camera signal starts at approximately $5 \times 10^7$ counts, coinciding with the onset of the sideview1 camera signal, indicating that certain shots, likely with a beam peak current just above the ionization threshold of hydrogen molecules, only generate plasma and excite wakes near the vacuum waist. But this density is not large enough to guide the beam up to the sideview1 camera where it could also produce ionization to the same degree and thereby producing light. It is not surprising that this translates to a large spread in the correlation plot shown in Fig. 5(b). Despite this spread it is clear that the sideview1 camera signal increases with the topview camera signal but then begins to saturate. The correlation implies that as the drive beam's peak current rises, it creates a larger wake that eventually dumps more energy per unit length along its path, resulting in stronger visible light emissions detected by both cameras. The saturation of the sideview1 signal, despite continued increases in the topview camera signal, suggests that the energy deposition at the sideview1 location has plateaued, possibly because the plasma and wake originated further upstream for beams with higher peak currents.

Fig. 5(c) illustrates the correlation between the signals from the sideview2 camera, positioned further downstream, and the sideview1 camera signal. For the shots yielding the lowest sideview1 signals, the sideview2 signal is nearly zero, indicating that the plasma and wake are present at the sideview1 location but do not extend to the more distant sideview2 location. As the drive beam intensity increases, signals from both sideview1 and sideview2 cameras increase, and a linear correlation emerges. However, for the shots with the highest sideview1 signals, the sideview2 camera's plasma light collection tends to saturate, corresponding with a shift of the entire plasma upstream, as discussed earlier. These inter-camera correlations suggest that in most cases, the plasma and wake extend across the entire 1.8-meter span from the topview camera to the downstream sideview2 camera. These are the longest (1.8m) beam-driven plasma wakes in a relatively high density ($3.5 \times 10^{16}$ cm$^{-3}$) plasma that the authors are aware of. We note that the datasets obtained at other pressures (e.g., 1.5 and 2.0 Torr) show similar results.

### 3.3 Visible plasma light vs. x-ray signal

Despite the facts that the x-ray diagnostic measures the longitudinally integrated betatron emission, whereas the topview and sideview cameras record the plasma light locally, we show in this section that these two observables are correlated and therefore, could provide additional information about the beam-plasma interaction.

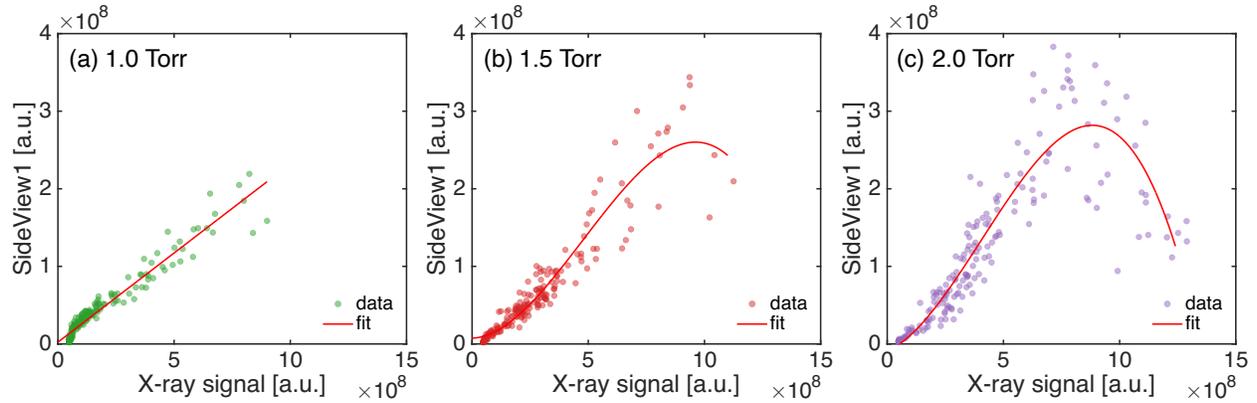

*Fig. 6. Correlation between plasma light and x-ray signal. The plasma light detected by the sideview1 camera is plotted as a function of the x-ray signal for three datasets acquired at different pressures. The red lines are polynomial fits [with order one (linear) for (a) and three (cubic) for (b) and (c)] to illustrate the trend of the data.*

In Fig. 6, the plasma light captured by the sideview1 camera is presented as a function of x-ray signal for three different gas pressures. For the 1.0 Torr dataset in Fig. 6(a), a notable linear correlation emerges initially between the visible plasma light and betatron x-ray signals. This correlation is particularly significant given the localized collection of plasma light (covering roughly 1.5 cm), in contrast to the meter-scale plasma length over which the x-ray radiation is generated. This linear correlation is notable considering that the plasma light is collected locally (spanning approximately 1.5 cm, much shorter than the meter-scale plasma length along the longitudinal direction). In the 1.5 Torr dataset depicted in Fig. 6(b), a similar pattern of correlation is observed, but the slope of the linear part is larger indicating that the electrons are losing more energy at the turning points of their betatron orbits due to a larger acceleration in higher plasma density wakes. Also, at the highest values of the x-ray signal there is some indication of saturation.

As the gas pressure further increases to 2.0 Torr, Fig. 6(c) reveals a distinct behavior: the visible plasma light initially increases linearly with the x-ray signal for shots with smaller x-ray signals, as anticipated. However, after peaking at an x-ray signal of $8 \times 10^8$, it declines almost linearly with a slope similar to the increase phase, despite the continued rise in x-ray signal. This trend could be attributed to a shift of the plasma upstream for shots with higher peak currents, resulting in plasma formation further upstream as explained earlier. Consequently, by the time the beam reaches the sideview1 camera location, it has been partially depleted and is less efficient in exciting wakes and depositing energy into the plasma. Alternatively, this pattern might indicate that the rear of the drive bunch enters the accelerating phase, extracting energy from the wake and thereby reducing the energy deposited into the plasma, which in turn lowers the plasma light yield. Notably, for datasets at pressures of 1.5 Torr and above, we observed using the imaging spectrometer that electrons had gained multiple GeV of energy [8].

These results indicate that, upon appropriate calibration, plasma light can be utilized as a noninvasive diagnostic tool. It is noteworthy that similar linear correlations between plasma light and energy deposition have recently been reported in other plasma wakefield acceleration (PWFA) experiments [29,30].

## 4 Conclusion

In this paper, we have documented correlations observed from the E300, plasma wakefield acceleration experiments at FACET-II which reveal the connections between maximum energy loss of the drive bunch, energy deposition into the plasma, betatron x-ray signal, and visible plasma light. These findings enable the estimation of energy dynamics of the drive bunch and plasma characteristics in scenarios where a direct measurement faces limitations, such as estimating the maximum energy loss of the drive bunch where the direct measurement using imaging spectrometer is unavailable, determining the length of plasma and wake and their upstream shift for bunches with higher peak currents. These correlations show the potential of x-ray and visible light emissions as noninvasive diagnostics for rapid sorting of large amounts of data acquired in a high repetition rate experiment and understanding the complicated energy transfer processes within plasma accelerators.

**Acknowledgements**

This work was supported at UCLA by the U.S. Department of Energy through Grant No. DE-SC0010064, and at SLAC by the U.S. Department of Energy under Contract Number DE-AC02-76SF00515. This research used resources of the National Energy Research Scientific Computing Center (NERSC), a U.S. Department of Energy Office of Science User Facility located at Lawrence Berkeley National Laboratory, operated under Contract No. DE-AC0205CH11231 using NERSC Award HEP-ERCAP-MP113. The authors thank Qianqian Su, Lance Hildebrand and Yujian Zhao for their help with the QPAD code development, as well as Lauren Alsberg, Max Gilljohann, Carsten Hast, Ryan Loney, Aime Matheron, and Marcellus Parker for their help with the experiment. A Knetsch and P San Miguel Claveria were supported by the France-Stanford Center for Interdisciplinary Studies for their travels to SLAC National Accelerator Laboratory. E Gerstmayr was supported by the U.S. Department of Energy, Office of Science, Fusion Energy Sciences under Award DE-SC0020076.